\documentclass[Afour,sageh,times]{sagej}

\usepackage{moreverb,url}
\usepackage{amsmath,amssymb,amsthm}
\usepackage{booktabs}
\usepackage{array}

\usepackage{enumitem}

\newtheorem{proposition}{Proposition}
\newtheorem{corollary}{Corollary}

\usepackage[colorlinks,bookmarksopen,bookmarksnumbered,citecolor=red,urlcolor=red]{hyperref}

\newcommand\BibTeX{{\rmfamily B\kern-.05em \textsc{i\kern-.025em b}\kern-.08em
T\kern-.1667em\lower.7ex\hbox{E}\kern-.125emX}}

\def\volumeyear{2016}

\begin{document}

\runninghead{Smith and Wittkopf}

\title{AGI and the Limits of Value Production}

\author{Zichen Song\affilnum{1,2}}

\affiliation{\affilnum{1}Department of Computer Science and Engineering, Sungkyunkwan University, South Korea\\
\affilnum{2}School of Information Science and Engineering, Lanzhou University, China}

\corrauth{Zichen Song, Department of Computer Science and Engineering, Sungkyunkwan University, South Korea; School of Information Science and Engineering, Lanzhou University, China.}

\email{sls530@skku.edu; songzichen894@gmail.com}

\begin{abstract}
This paper develops a political-economy model of artificial general intelligence (AGI) as a technology that progressively substitutes living labor with machine-based productive systems. The model studies the transition from the first moment at which AGI becomes economically capable of replacing labor to the later moment at which AGI becomes technically and actually capable of near-complete replacement. The central distinction is between technical substitutability and actual adoption. Technical substitutability is the feasible replacement ceiling implied by the state of AGI capability, whereas actual adoption is the realized replacement share chosen under cost, profitability, and adoption frictions. Under the strict value-theoretic assumption that AGI transfers value but does not itself create new value, deeper AGI adoption raises the organic composition of capital, reduces the quantity of living labor when adoption outpaces the creation of new labor fields, compresses the source of surplus value, and places downward pressure on the social rate of profit. In the limiting case in which actual AGI adoption approaches complete substitution and new labor fields fail to compensate for displaced labor, living labor tends to zero, surplus value tends to zero, and the profit rate tends to zero. The model therefore identifies near-complete AGI substitution not merely as an efficiency transition, but as a boundary case for value production under a strict political-economy theory of value.
\end{abstract}


\maketitle

\section{Introduction}

Artificial general intelligence raises a theoretical problem that differs from ordinary technological change. Earlier waves of mechanization and automation replaced particular tasks, reorganized labor processes, and increased productivity, but they generally left large domains of living labor intact or created new labor fields into which workers could be reabsorbed. AGI is distinctive because it may expand the frontier of substitution beyond routine manual or cognitive tasks toward a much wider range of productive, administrative, analytical, and coordinative activities. The relevant question is therefore not only whether AGI increases efficiency, but whether it can push the structure of production from partial labor substitution toward near-complete labor substitution.

This paper develops a political-economy model of that transition. The central object is the movement from the first moment at which AGI becomes economically capable of replacing labor to the later moment at which AGI becomes technically and actually capable of near-complete replacement. The model distinguishes two concepts that are often conflated. The first is technical substitutability: the share of labor tasks that AGI can in principle perform at a given level of capability. The second is actual adoption: the share of labor tasks that capital actually replaces after accounting for cost, profitability, and adoption frictions. This distinction matters because a task may become technically automatable before it is economically attractive or organizationally adopted. The paper therefore models technical substitutability as a ceiling, while actual adoption follows a continuous adjustment process toward that ceiling.

The model is built on a strict political-economy value assumption: living labor creates new value, while AGI-based systems transfer, reorganize, and cheapen production without themselves creating new value. Under this assumption, AGI substitution changes the internal composition of capital. As actual adoption rises, the amount of living labor employed in production falls, variable capital declines, and AGI-related constant capital rises. The organic composition of capital therefore increases with actual AGI adoption. This is not a merely descriptive claim about technological intensity; it is a structural claim about the changing relation between machine-based production systems and living labor within the process of value production.

The key mechanism is straightforward. Let AGI capability grow over time. As capability rises, the technical substitution ceiling increases, while the unit cost of AGI replacement falls. Once replacement cost falls to the level of wage labor, capital has an economic incentive to accelerate adoption. Actual adoption does not jump immediately to the technical ceiling; it adjusts continuously, governed by the gap between what AGI can technically replace and what capital has actually adopted. As actual adoption deepens, the remaining quantity of living labor depends on whether new labor fields expand fast enough to offset the labor displaced by AGI. If new labor fields grow sufficiently quickly, the system may continue to reproduce living labor in new forms. If actual substitution outpaces the expansion of new labor fields, living labor declines.

This distinction allows the paper to identify when AGI becomes structurally destabilizing for value production: not merely when it replaces old tasks, but when actual AGI adoption absorbs existing labor tasks faster than new labor fields are created. In that case, the total quantity of living labor falls, surplus value contracts with it under the strict value-theoretic formulation, and AGI adoption simultaneously raises constant capital relative to variable capital. The model therefore predicts rising organic composition and downward pressure on the social profit rate when substitution is not offset by new labor domains. The limiting result is sharper: if actual AGI adoption approaches complete substitution while new labor fields fail to compensate for displaced living labor, then living labor, surplus value, and the social profit rate all tend toward zero. Near-complete AGI substitution is therefore not modeled as a simple endpoint of technological efficiency, but as a boundary case in which a system organized around value expansion confronts the erosion of the living-labor basis of value production.

The contribution of this paper is threefold:

\begin{itemize}
    \item This paper formalizes the transition from partial to near-complete AGI substitution as a dynamic political-economy process rather than a static comparison between human labor and machine production. The model identifies the movement from the initial economic feasibility of substitution to the later stage of near-complete replacement.

    \item This paper separates technical substitutability from actual adoption. Technical substitutability defines what AGI can in principle replace, while actual adoption captures what capital actually replaces under cost, profitability, and adoption frictions. This distinction allows the model to derive separate threshold times for near-complete technical feasibility and near-complete actual adoption.

    \item This paper shows that, under a strict political-economy theory of value, deeper AGI adoption monotonically raises the organic composition of capital and places downward pressure on the social profit rate. If actual AGI adoption persistently outpaces the creation of new labor fields, the system moves toward a limiting case in which living labor, surplus value, and the profit rate tend toward zero.
\end{itemize}

\section{Model Setup}

This section introduces the minimal state variables required to describe the transition from partial AGI labor substitution to near-complete AGI labor substitution. The purpose of the setup is not to model all institutional features of capitalist production, but to isolate the core dynamic relation between AGI capability, actual adoption, and the remaining quantity of living labor. The model therefore begins with time, the level of AGI capability, and the actual rate at which labor tasks are replaced by AGI-based systems.

Let time be denoted by \(t\), with
\begin{equation}
t\geq 0.
\end{equation}
The initial date \(t=0\) should be interpreted as the beginning of the modeled transition rather than the historical beginning of automation. All subsequent variables are evaluated along this transition path. This convention allows the model to distinguish whether a given substitution threshold lies in the future, occurs at the initial date, or has already occurred before the modeled period begins.

Let the level of AGI capability be denoted by \(A(t)\). AGI capability summarizes the general productive, cognitive, coordinative, and technical power of AGI-based systems at time \(t\). It is not itself the realized replacement of labor. Rather, it determines the technological frontier within which replacement may become feasible. AGI capability is assumed to grow over time:
\begin{equation}
A'(t)>0.
\end{equation}
This assumption captures the idea that AGI systems become progressively more capable during the transition, reflecting improvements in model architecture, reasoning ability, tool use, data access, software integration, inference cost, robotics, or organizational deployment capacity. The model does not require specifying which of these channels dominates; it only requires that the overall capability index rises over time. The representative transition studied in this paper is the interval between a partial-substitution threshold and a near-complete-substitution threshold, where the partial-substitution threshold marks the point at which AGI first becomes economically relevant as a substitute for labor tasks, and the near-complete-substitution threshold marks the point at which AGI approaches the ability, and eventually the actual adoption level, required to replace almost all labor tasks covered by the model. The key state variable is adoption rate of AGI, denoted by \(\alpha(t)\), with
\begin{equation}
\alpha(t)\in[0,1].
\end{equation}
This variable measures realized replacement, not merely technical feasibility. When \(\alpha(t)=0\), no labor tasks are actually replaced by AGI in the modeled production system. When \(0<\alpha(t)<1\), AGI partially replaces labor, so production still depends on a positive mass of living labor. When \(\alpha(t)\) approaches one, AGI approaches near-complete actual replacement, and the amount of living labor directly employed in the modeled production process becomes correspondingly small. This distinction between actual adoption and technical feasibility is essential for the rest of the paper, because AGI may be technically capable of replacing a task before capital actually adopts it under cost, profitability, and adjustment frictions.

\section{Substitutability and Cost}

The first component of the model is the technical substitution ceiling, which represents the maximum share of labor tasks that artificial general intelligence (AGI) can perform at a given capability level. Let \(q(A)\) denote this share at capability level \(A\). In line with economic reasoning, we assume that the capacity for AGI to replace tasks grows with its capability, but at a decreasing marginal rate, reflecting technological limits and increasing complexity of remaining tasks:
\begin{equation}
q'(A)>0,
\end{equation}
\begin{equation}
q''(A)<0.
\end{equation}
A tractable functional form that satisfies these properties is
\begin{equation}
q(A)=1-e^{-\beta A},
\end{equation}
where
\begin{equation}
\beta>0.
\end{equation}
Differentiating this function yields
\begin{equation}
q'(A)=\beta e^{-\beta A},
\end{equation}
which is strictly positive for all \(A>0\), confirming that technical substitutability increases with AGI capability. The second derivative is
\begin{equation}
q''(A)=-\beta^2 e^{-\beta A},
\end{equation}
which is strictly negative, illustrating that the marginal increase in technical substitutability diminishes as AGI becomes more capable. The limiting value of the technical ceiling is
\begin{equation}
\lim_{A\to\infty}q(A)=1,
\end{equation}
indicating that, at sufficiently high levels of AGI capability, virtually all labor tasks could theoretically be performed by machines.

The second component is the unit cost of AGI replacement, denoted by \(m(A)\), which measures the capital expenditure required to substitute one unit of labor with AGI, including costs for software, hardware, deployment, monitoring, maintenance, and integration. The model assumes that unit replacement costs decline as AGI capability rises:
\begin{equation}
m'(A)<0.
\end{equation}
A simple functional form capturing this decline is
\begin{equation}
m(A)=m_0 e^{-\gamma A},
\end{equation}
where
\begin{equation}
m_0>0,
\end{equation}
and
\begin{equation}
\gamma>0.
\end{equation}
Differentiating the cost function gives
\begin{equation}
m'(A)=-\gamma m_0 e^{-\gamma A},
\end{equation}
which is strictly negative for all \(A>0\), confirming that more capable AGI reduces the per-task replacement cost. This decreasing cost function captures the political-economic insight that technological advances enable capital to economize on living labor more effectively over time, setting the stage for gradual or accelerated adoption of automation.

Together, these two components—technical substitutability and declining unit replacement cost—establish the foundational conditions for analyzing the dynamic adoption of AGI. They define both the ceiling of what is technologically possible and the economic incentives that determine when and how quickly capital will choose to implement labor substitution. These parameters are crucial for later deriving the thresholds at which partial substitution becomes feasible and near-complete substitution is approached.

\section{Value, Variable Capital, and Surplus Value}

The third component of the model is the value structure of production. This section specifies how living labor, variable capital, and surplus value are represented before AGI-related constant capital is introduced in the next section. The purpose is to isolate the strict value-theoretic assumption used throughout the paper: new value is generated by living labor, while AGI-based systems can reorganize, cheapen, or accelerate production without themselves serving as an independent source of new value. The adoption rate \(\alpha\) therefore does not merely describe a technical replacement ratio. It also determines how much living labor remains within production and, through that channel, how much variable capital and surplus value remain in the system.

Let \(L\) denote the total scale of labor tasks. If the actual adoption rate is \(\alpha\), then the amount of living labor still employed in production is
\begin{equation}
H(\alpha)=(1-\alpha)L.
\end{equation}
This expression states that living labor is the residual task mass not yet replaced by AGI. When \(\alpha=0\), all labor tasks remain performed by living labor. When \(\alpha\) rises, the living-labor component contracts proportionally. When \(\alpha\) approaches one, the living-labor component approaches zero, provided the total task scale \(L\) is held fixed.

The quantity of labor tasks actually replaced by AGI is
\begin{equation}
\alpha L.
\end{equation}
Thus the task space is divided into two parts: the still-living-labor component \((1-\alpha)L\) and the AGI-replaced component \(\alpha L\). This decomposition is important because the model attributes new value only to the former. The replaced component may still produce outputs, reduce costs, or raise technical efficiency, but under the strict political-economy value assumption used here, it does not directly generate new value.

Let \(\varphi\) denote the new value created by one unit of living labor. Then total new value created by living labor is
\begin{equation}
Y_v=\varphi(1-\alpha)L.
\end{equation}
This equation links value creation to the remaining quantity of living labor. The parameter \(\varphi\) measures the value-creating capacity of each unit of living labor, while \((1-\alpha)L\) measures how much living labor remains active in production. Hence AGI adoption affects total new value through the reduction of the living-labor base.

Let \(w\) denote the wage paid per unit of living labor. Variable capital is therefore
\begin{equation}
v=w(1-\alpha)L.
\end{equation}
Variable capital is the wage bill advanced to living labor. Since only the unreplaced portion of labor receives wages in this simplified structure, variable capital falls mechanically as actual AGI adoption rises. This is the first channel through which substitution changes the structure of capital: it reduces the wage-mediated component of production.

Surplus value is new value minus wages:
\begin{equation}
s=Y_v-v.
\end{equation}
This definition separates the total new value created by living labor from the portion returned to labor as wages. The residual is surplus value. In this formulation, the source of surplus value is not the AGI system itself, but the difference between the value produced by living labor and the wage cost of that living labor.

Substituting the expressions for \(Y_v\) and \(v\) yields
\begin{equation}
s=\varphi(1-\alpha)L-w(1-\alpha)L.
\end{equation}
Both terms contain the same living-labor base \((1-\alpha)L\). The first term is the gross new value created by living labor, and the second term is the wage cost paid to that living labor. AGI adoption reduces both terms by reducing the quantity of living labor that remains in production.

Hence
\begin{equation}
s=(\varphi-w)(1-\alpha)L.
\end{equation}
This expression shows that surplus value depends on two components: the per-unit gap between value created and wage paid, \(\varphi-w\), and the remaining quantity of living labor, \((1-\alpha)L\). Even if the per-unit surplus remains positive, total surplus value contracts when the living-labor base contracts.

For positive surplus value, the model requires
\begin{equation}
\varphi>w.
\end{equation}
This condition states that each unit of living labor must create more new value than it receives as wages. If this inequality fails, surplus value is non-positive and the value-expansion mechanism analyzed in the paper does not operate. The condition is therefore not a behavioral assumption about workers or firms, but a consistency requirement for positive surplus value within this value structure.

Define the surplus-value rate by
\begin{equation}
e=\frac{s}{v}.
\end{equation}
The surplus-value rate measures surplus value relative to variable capital. It summarizes how much surplus is generated per unit of wage expenditure. In the present model, this rate is used to rewrite surplus value in a compact form that will later enter the profit-rate expression.

Using the expressions above gives
\begin{equation}
e=\frac{\varphi-w}{w}.
\end{equation}
This follows because both surplus value and variable capital are proportional to the same living-labor quantity \((1-\alpha)L\). The adoption rate cancels out in the ratio, so the surplus-value rate depends on the per-unit value-wage relation rather than on the scale of living labor itself. Actual AGI adoption changes the mass of surplus value by changing the amount of living labor, but it does not by itself change \(e\) in this baseline specification.

Equivalently, surplus value can be written as
\begin{equation}
s=ev.
\end{equation}
This identity expresses surplus value as the surplus-value rate multiplied by variable capital. It is useful because it makes the later dynamic implication transparent: if variable capital falls as actual AGI adoption rises, then surplus value falls proportionally unless the surplus-value rate rises enough to offset that decline.

Since
\begin{equation}
v=w(1-\alpha)L,
\end{equation}
the surplus-value expression becomes
\begin{equation}
s(\alpha)=ew(1-\alpha)L.
\end{equation}

This final expression is the central value equation of the model. It says that, under the strict political-economy value assumption, the mass of surplus value is proportional to the quantity of living labor that remains in production. A rise in actual AGI adoption reduces \((1-\alpha)L\), thereby reducing variable capital and surplus value, unless some other force such as a higher surplus-value rate, a higher wage-normalized value product, or an expansion of the total labor-task field offsets the decline. AGI may cheapen production, transfer value, reorganize labor processes, and increase technical output, but in this strict formulation it does not itself create new value. This is why the later sections focus on whether new labor fields expand fast enough to prevent the living-labor base from shrinking as actual AGI adoption approaches near-complete substitution.

\section{Capital Structure and Organic Composition}

The fourth component of the model concerns the structure of capital and how it interacts with AGI-driven labor substitution. Pre-existing constant capital, denoted \(c_0\), represents the investment in machinery, infrastructure, and other fixed productive assets that is independent of labor hours. As AGI adoption proceeds, additional constant capital is introduced to replace \(\alpha L\) labor tasks, where \(\alpha\) is the actual adoption rate and \(L\) is the total labor task scale. The unit AGI replacement cost, \(m(A)\), reflects the capital expenditure required per replaced task, encompassing model deployment, computational infrastructure, software, and maintenance. Therefore, the AGI-related constant capital is

\begin{equation}
c_A=m(A)\alpha L.
\end{equation}

Total constant capital is the sum of pre-existing and AGI-related capital:

\begin{equation}
c(\alpha,A)=c_0+m(A)\alpha L.
\end{equation}

Using the explicit exponential cost function that declines with AGI capability,

\begin{equation}
c(\alpha,A)=c_0+m_0e^{-\gamma A}\alpha L,
\end{equation}

where \(m_0\) is the initial replacement cost and \(\gamma\) controls the rate at which costs decline as AGI capability \(A\) grows.

Total advanced capital, \(K(\alpha,A)\), combines constant and variable capital, the latter being the wage cost of the remaining living labor:

\begin{equation}
K(\alpha,A)=c(\alpha,A)+v(\alpha),
\end{equation}
\begin{equation}
K(\alpha,A)=c_0+m(A)\alpha L+w(1-\alpha)L,
\end{equation}
\begin{equation}
K(\alpha,A)=c_0+m_0e^{-\gamma A}\alpha L+w(1-\alpha)L.
\end{equation}

The organic composition of capital, \(\Omega(\alpha,A)\), is defined as the ratio of constant to variable capital:

\begin{equation}
\Omega(\alpha,A)=\frac{c(\alpha,A)}{v(\alpha)},
\end{equation}
\begin{equation}
\Omega(\alpha,A)=\frac{c_0+m(A)\alpha L}{w(1-\alpha)L},
\end{equation}
\begin{equation}
\Omega(\alpha,A)=\frac{c_0+m_0e^{-\gamma A}\alpha L}{w(1-\alpha)L}.
\end{equation}

As actual AGI adoption \(\alpha\) approaches complete replacement, the variable capital component diminishes:

\begin{equation}
\lim_{\alpha\to 1^-}w(1-\alpha)L=0,
\end{equation}

while the numerator remains positive provided \(c_0+m(A)L>0\), implying

\begin{equation}
\lim_{\alpha\to 1^-}\Omega(\alpha,A)=+\infty.
\end{equation}

This demonstrates that as AGI adoption deepens, the production process becomes increasingly capital-intensive relative to remaining living labor.

To confirm monotonicity, differentiate \(\Omega(\alpha,A)\) with respect to \(\alpha\), holding \(A\) fixed:

\begin{equation}
\frac{\partial \Omega}{\partial \alpha}=
\frac{\partial}{\partial\alpha}\left[\frac{c_0+m(A)\alpha L}{w(1-\alpha)L}\right].
\end{equation}

The expression is a ratio. The numerator increases with \(\alpha\), because additional AGI adoption adds constant capital. The denominator decreases with \(\alpha\), because higher adoption reduces the remaining wage-paying labor share. Applying the quotient rule gives:

\begin{equation}
\frac{\partial \Omega}{\partial \alpha}=
\frac{m(A)L\cdot w(1-\alpha)L-(c_0+m(A)\alpha L)(-wL)}{[w(1-\alpha)L]^2}.
\end{equation}

The first term in the numerator comes from differentiating constant capital, while the second term comes from differentiating variable capital in the denominator. Since the derivative of \(w(1-\alpha)L\) is \(-wL\), the second component becomes positive after subtracting a negative term. The numerator can therefore be rewritten as:

\begin{equation}
\begin{aligned}
&m(A) w L^2 (1-\alpha) + w L (c_0 + m(A) \alpha L) \\
&= w L \bigl[ m(A)L(1-\alpha) + c_0 + m(A)\alpha L \bigr].
\end{aligned}
\end{equation}

Inside the bracket, the two terms involving \(m(A)\) combine because one corresponds to the non-adopted share and the other to the adopted share:

\begin{equation}
m(A)L(1-\alpha)+m(A)\alpha L=m(A)L.
\end{equation}

Thus the numerator collapses to \(wL[c_0+m(A)L]\), and the derivative becomes:

\begin{equation}
\frac{\partial \Omega}{\partial \alpha}=\frac{wL[c_0+m(A)L]}{w^2(1-\alpha)^2L^2}.
\end{equation}

After cancelling the common factor \(wL\), we obtain the simplified derivative:

\begin{equation}
\frac{\partial \Omega}{\partial \alpha}=\frac{c_0+m(A)L}{wL(1-\alpha)^2}.
\end{equation}

The sign of this derivative follows directly from the maintained assumptions:

\begin{equation}
\left.
\begin{cases}
c_0>0,\\
m(A)>0,\\
w>0,\\
L>0,\\
0\leq \alpha<1
\end{cases}
\right\}
\text{Assumptions for positivity.}
\end{equation}

Under these conditions, the numerator is positive and the denominator is also positive. Therefore,

\begin{equation}
\frac{\partial \Omega}{\partial \alpha}>0.
\end{equation}

This formal derivation confirms that the organic composition rises monotonically with actual AGI adoption. Intuitively, each additional unit of labor replaced by AGI increases constant capital while simultaneously reducing variable capital, thereby structurally intensifying the capital-labor composition in production. The monotonic increase of \(\Omega(\alpha,A)\) underscores the structural implications of AGI substitution: the deeper the adoption, the more the production process relies on machine-based systems relative to living labor, reinforcing the predicted downward pressure on the social profit rate and highlighting the long-term tension between automation and the maintenance of value-creating labor inputs.

\section{Profit Rate and the Limit of Complete Substitution}

The social profit rate measures the relation between surplus value and the total capital advanced in production. In the present model, total advanced capital consists of constant capital, which includes pre-existing capital and AGI-related capital outlays, and variable capital, which is the wage bill paid to living labor. The social profit rate is therefore defined as surplus value divided by total advanced capital:
\begin{equation}
r(\alpha,A)=\frac{s(\alpha)}{c(\alpha,A)+v(\alpha)}.
\end{equation}

This definition now allows the previous components of the model to be combined. Surplus value is generated by living labor and is therefore proportional to the remaining non-substituted labor share. Constant capital increases with AGI adoption, while variable capital decreases as living labor is replaced. Substituting the expressions above gives
\begin{equation}
r(\alpha,A)=
\frac{
ew(1-\alpha)L
}{
c_0+m(A)\alpha L+w(1-\alpha)L
}.
\end{equation}

The denominator contains three distinct components. The term \(c_0\) is pre-existing constant capital, \(m(A)\alpha L\) is the AGI-related constant capital associated with actual substitution, and \(w(1-\alpha)L\) is the remaining variable capital. If the explicit replacement-cost function is used, the same profit-rate expression becomes
\begin{equation}
r(\alpha,A)=
\frac{
ew(1-\alpha)L
}{
c_0+m_0e^{-\gamma A}\alpha L+w(1-\alpha)L
}.
\end{equation}

The central comparative-static question is whether deeper actual AGI adoption raises or lowers the social profit rate in this strict value-theoretic setting. The answer is not immediate from visual inspection, because increasing \(\alpha\) has two opposing appearances: it reduces the wage bill, but it also reduces the living-labor basis of surplus value and increases the relative role of constant capital. To evaluate the net effect, define the numerator and denominator of the profit rate as
\begin{equation}
N(\alpha)=ew(1-\alpha)L,
\end{equation}
and
\begin{equation}
D(\alpha)=c_0+m(A)\alpha L+w(1-\alpha)L.
\end{equation}

With this notation, the profit rate can be written in the compact form
\begin{equation}
r(\alpha,A)=\frac{N(\alpha)}{D(\alpha)}.
\end{equation}

The numerator declines with actual AGI adoption because a higher adoption rate reduces the amount of living labor that remains in production. Its derivative is
\begin{equation}
N'(\alpha)=-ewL,
\end{equation}
and
\begin{equation}
D'(\alpha)=m(A)L-wL.
\end{equation}
Equivalently,
\begin{equation}
D'(\alpha)=\left[m(A)-w\right]L.
\end{equation}

The sign of \(D'(\alpha)\) depends on whether AGI replacement cost is above or below the wage. However, the sign of the profit-rate derivative is not determined by \(D'(\alpha)\) alone, because the numerator also changes. Applying the quotient rule gives
\begin{equation}
\frac{\partial r}{\partial \alpha}
=
\frac{N'(\alpha)D(\alpha)-N(\alpha)D'(\alpha)}{D(\alpha)^2}.
\end{equation}

Substituting the expressions for \(N'(\alpha)\), \(N(\alpha)\), and \(D'(\alpha)\) gives the full derivative before simplification:
\begin{equation}
\frac{\partial r}{\partial \alpha}
=
\frac{
(-ewL)D(\alpha)
-
ew(1-\alpha)L\left[m(A)-w\right]L
}{
D(\alpha)^2
}.
\end{equation}

The next step is to isolate the common negative factor. This makes clear that the derivative will be negative if the remaining bracketed term is positive. Factoring out \(-ewL\) gives
\begin{equation}
\frac{\partial r}{\partial \alpha}
=
-
\frac{
ewL
\left[
D(\alpha)+(1-\alpha)L\left(m(A)-w\right)
\right]
}{
D(\alpha)^2
}.
\end{equation}

It remains to simplify the bracket. The bracket combines the original denominator with the correction term produced by the fact that variable capital and AGI-related constant capital move in opposite directions as \(\alpha\) changes. Expanding the bracket gives
\begin{equation}
\begin{aligned}
&D(\alpha) + (1-\alpha)L\bigl(m(A)-w\bigr) \\
&= c_0 + m(A)\alpha L \\
&\quad + w(1-\alpha)L \\
&\quad + (1-\alpha)Lm(A) - (1-\alpha)Lw .
\end{aligned}
\end{equation}

The two wage terms cancel exactly:
\begin{equation}
w(1-\alpha)L-(1-\alpha)Lw=0.
\end{equation}

After this cancellation, the expression reduces to the pre-existing constant capital plus the AGI-capital terms:
\begin{equation}
c_0+m(A)\alpha L+(1-\alpha)Lm(A).
\end{equation}

The AGI-capital terms can then be combined because they contain the same unit replacement cost \(m(A)\) and the same task scale \(L\):
\begin{equation}
c_0+m(A)L\left[\alpha+1-\alpha\right].
\end{equation}

Since the adoption shares inside the bracket sum to one, the full bracket collapses to a strictly positive capital term:
\begin{equation}
D(\alpha)+(1-\alpha)L\left(m(A)-w\right)
=
c_0+m(A)L.
\end{equation}

Therefore, the derivative of the social profit rate with respect to actual AGI adoption is
\begin{equation}
\frac{\partial r}{\partial \alpha}
=
-
\frac{
ewL\left[c_0+m(A)L\right]
}{
\left[c_0+m(A)\alpha L+w(1-\alpha)L\right]^2
}.
\end{equation}

This expression has a clear sign under the maintained assumptions of the model. The numerator inside the fraction is positive, the denominator is squared and therefore positive, and the whole expression is preceded by a negative sign. Under
\begin{equation}
e>0,\quad w>0,\quad L>0,\quad c_0>0,\quad m(A)>0,
\end{equation}
this derivative is strictly negative:
\begin{equation}
\frac{\partial r}{\partial \alpha}<0.
\end{equation}

The result is not driven by demand spillovers or price competition. It follows directly from the strict value-theoretic premise that living labor creates surplus value while AGI raises the constant-capital share of production. Even when AGI becomes cheaper than wage labor, deeper actual substitution reduces the living-labor basis of surplus value. In this formulation, the wage saving associated with substitution does not overturn the fact that the source of surplus value contracts as living labor contracts.

The same logic also determines the limiting case of complete substitution. As actual adoption approaches complete substitution, the remaining living-labor share approaches zero:
\begin{equation}
\lim_{\alpha\to 1^-}(1-\alpha)L=0.
\end{equation}

Since surplus value is proportional to living labor in the model, the disappearance of living labor implies the disappearance of surplus value:
\begin{equation}
\lim_{\alpha\to 1^-}s(\alpha)=0.
\end{equation}

The denominator of the profit rate does not vanish in the same limit. Variable capital tends to zero, but constant capital remains positive because pre-existing capital and AGI-related capital remain. The denominator of the profit rate approaches
\begin{equation}
\lim_{\alpha\to 1^-}\left[c_0+m(A)\alpha L+w(1-\alpha)L\right]
=
c_0+m(A)L.
\end{equation}

The limiting profit rate is therefore zero:
\begin{equation}
\lim_{\alpha\to 1^-}r(\alpha,A)=0.
\end{equation}

This is the model's boundary result: if actual AGI adoption tends to complete substitution and living labor tends to zero, then surplus value tends to zero and the social profit rate tends to zero under the strict political-economy value assumption. Complete substitution is therefore not simply a limiting case of technological efficiency. In this framework, it is also the limiting case in which the living-labor basis of value production is exhausted while constant capital remains positive.

\section{Dynamic Transition and Threshold Times}

We now introduce explicit time dynamics in order to describe the movement from the initial economic feasibility of AGI substitution to the later stage of near-complete technical substitutability. The purpose of this section is to specify how AGI capability evolves, how replacement cost changes with that capability, and how the technical ceiling of substitution moves over time. Let AGI capability evolve as
\begin{equation}
A(t)=A_0e^{gt},
\end{equation}
where
\begin{equation}
A_0>0,\quad g>0.
\end{equation}
The parameter \(A_0\) denotes the initial level of AGI capability, while \(g\) denotes the growth rate of that capability. This specification captures the idea that AGI capability expands continuously over time rather than arriving as a one-time discrete shock.

Given this capability path, the unit cost of replacing one labor task with AGI also becomes time-dependent. Since the model assumes that higher AGI capability reduces the unit replacement cost, the cost path is obtained by evaluating the replacement-cost function along the capability trajectory:
\begin{equation}
m(t)=m_0e^{-\gamma A(t)}.
\end{equation}
Substituting the trajectory of \(A(t)\) gives
\begin{equation}
m(t)=m_0e^{-\gamma A_0e^{gt}}.
\end{equation}
Thus, as AGI capability rises, the unit cost of replacement falls. This declining cost path is the economic basis for the later emergence of a substitution threshold: even if AGI is technically improving, large-scale adoption becomes economically relevant only when replacement cost becomes sufficiently low relative to wage labor.

The technical substitution ceiling also evolves over time. This ceiling represents the maximum share of labor tasks that AGI can technically perform at a given moment. It is not yet the actual adoption rate; it is only the feasible upper bound implied by the state of AGI capability. The technical substitution ceiling over time is
\begin{equation}
q(t)=1-e^{-\beta A(t)}.
\end{equation}
Substituting \(A(t)\) gives
\begin{equation}
q(t)=1-e^{-\beta A_0e^{gt}}.
\end{equation}
This expression says that as AGI capability grows, the technical frontier of substitutable labor tasks expands toward one. However, the model does not assume that capital immediately adopts all technically feasible substitution.

It is important that \(q(t)\) is not the actual adoption rate. The variable \(q(t)\) is the technological ceiling: the share of labor tasks AGI is technically capable of replacing. The actual adoption rate \(\alpha(t)\) is chosen and realized through economic and organizational adoption dynamics. The model therefore imposes
\begin{equation}
0\leq \alpha(t)\leq q(t)\leq 1.
\end{equation}
This inequality summarizes the basic timing structure of the model. AGI may become technically capable of replacing a task before capital actually replaces that task. Actual adoption may lag behind technical feasibility because of cost, profitability conditions, organizational integration, and adoption frictions.

The first threshold is the cost threshold at which partial substitution becomes economically feasible. This is the point at which the unit AGI replacement cost falls to the level of the wage. Before this point, AGI may already possess some technical substitutability, but replacement is not yet cost-competitive with living labor. Let \(t_p\) be the time at which the AGI unit replacement cost equals the wage:
\begin{equation}
w=m(t_p).
\end{equation}
Using the cost path gives
\begin{equation}
w=m_0e^{-\gamma A(t_p)}.
\end{equation}
This equation defines the moment when the economic comparison between AGI replacement and wage labor becomes exact. To solve for that moment, take logarithms on both sides:
\begin{equation}
\ln w=\ln m_0-\gamma A(t_p).
\end{equation}
Rearranging gives the AGI capability level required for the unit replacement cost to equal the wage:
\begin{equation}
A(t_p)=\frac{1}{\gamma}\ln\frac{m_0}{w}.
\end{equation}
This expression identifies the capability threshold associated with the beginning of economically feasible substitution.

Since \(A(t)=A_0e^{gt}\), the capability threshold can be translated into a time threshold. Substituting the capability path into the previous condition gives
\begin{equation}
A_0e^{gt_p}=\frac{1}{\gamma}\ln\frac{m_0}{w}.
\end{equation}
Solving for \(t_p\) yields
\begin{equation}
t_p=
\frac{1}{g}
\ln\left[
\frac{1}{\gamma A_0}
\ln\frac{m_0}{w}
\right].
\end{equation}
This is the partial-substitution threshold time. It is the first time at which AGI replacement becomes economically competitive with wage labor under the cost path specified above.

This expression has existence and timing conditions. A real positive cost threshold requires
\begin{equation}
m_0>w,
\end{equation}
so that
\begin{equation}
\ln\frac{m_0}{w}>0.
\end{equation}
This condition means that, at the initial cost scale, AGI replacement begins above the wage level. If this condition failed, the cost threshold would not describe a future transition from expensive AGI to cost-competitive AGI in the intended sense.

For the threshold to occur weakly after the initial date \(t=0\), the following condition is required:
\begin{equation}
\ln\frac{m_0}{w}\geq \gamma A_0.
\end{equation}
This condition ensures that the capability needed to make AGI replacement cost equal to the wage is not already below the initial capability level. If the inequality is strict, then \(t_p>0\):
\begin{equation}
\ln\frac{m_0}{w}>\gamma A_0.
\end{equation}
If equality holds, then
\begin{equation}
t_p=0.
\end{equation}
In that boundary case, the model begins exactly at the moment when AGI replacement cost equals the wage. If
\begin{equation}
0<\ln\frac{m_0}{w}<\gamma A_0,
\end{equation}
then the threshold occurred before the model's initial date.
This last case means that AGI replacement was already economically competitive at \(t=0\), so the model would begin after the partial-substitution threshold rather than before it.

The second threshold is the near-complete technical substitution threshold. This threshold concerns technical feasibility, not actual adoption. It asks when the technical ceiling becomes sufficiently close to full substitutability. Let \(\bar{\alpha}\) denote a near-complete substitution benchmark, with
\begin{equation}
0<\bar{\alpha}<1.
\end{equation}
For example, \(\bar{\alpha}=0.99\) represents ninety-nine percent technical substitutability. Let \(t_f^{tech}\) be the time at which the technical ceiling reaches \(\bar{\alpha}\):
\begin{equation}
q(t_f^{tech})=\bar{\alpha}.
\end{equation}
This equation defines the time at which AGI is technically capable of replacing the benchmark share of labor tasks. Using the expression for \(q(t)\),
\begin{equation}
1-e^{-\beta A(t_f^{tech})}=\bar{\alpha}.
\end{equation}
The equation can be solved by first isolating the exponential term:
\begin{equation}
e^{-\beta A(t_f^{tech})}=1-\bar{\alpha}.
\end{equation}
Taking logarithms gives
\begin{equation}
-\beta A(t_f^{tech})=\ln(1-\bar{\alpha}).
\end{equation}
Therefore,
\begin{equation}
A(t_f^{tech})=-\frac{1}{\beta}\ln(1-\bar{\alpha}).
\end{equation}
This is the capability level required for the technical substitution ceiling to reach the near-complete benchmark.

Using \(A(t)=A_0e^{gt}\), this capability condition becomes
\begin{equation}
A_0e^{gt_f^{tech}}
=
-\frac{1}{\beta}\ln(1-\bar{\alpha}).
\end{equation}
Solving for \(t_f^{tech}\) gives
\begin{equation}
t_f^{tech}
=
\frac{1}{g}
\ln\left[
\frac{-\ln(1-\bar{\alpha})}{\beta A_0}
\right].
\end{equation}
This expression gives the near-complete technical substitution time: the moment when the technological ceiling reaches the benchmark \(\bar{\alpha}\). It does not yet imply that capital has actually adopted AGI to the same extent.

This threshold requires
\begin{equation}
0<\bar{\alpha}<1.
\end{equation}
The lower bound ensures that the benchmark is positive, while the upper bound ensures that the logarithm is well-defined and finite. For the threshold to occur weakly after the initial date, it must satisfy
\begin{equation}
-\ln(1-\bar{\alpha})\geq \beta A_0.
\end{equation}
Equivalently,
\begin{equation}
\bar{\alpha}\geq 1-e^{-\beta A_0}.
\end{equation}
Since
\begin{equation}
q(0)=1-e^{-\beta A_0},
\end{equation}
the condition can be written as
\begin{equation}
\bar{\alpha}\geq q(0).
\end{equation}
This says that the benchmark must not already be below the initial technical ceiling. If it were below the initial ceiling, then the near-complete technical threshold would have occurred before the model's starting date.

The threshold is strictly in the future when
\begin{equation}
-\ln(1-\bar{\alpha})>\beta A_0.
\end{equation}
Equivalently,
\begin{equation}
\bar{\alpha}>1-e^{-\beta A_0}.
\end{equation}
Thus, \(t_f^{tech}\) is a future technical threshold only when the near-complete benchmark lies strictly above the initial technical substitution ceiling. This completes the derivation of the two threshold times: \(t_p\), the economic cost threshold for partial substitution, and \(t_f^{tech}\), the technical threshold for near-complete substitutability.

\section{Continuous Actual Adoption and Living Labor Dynamics}

The model specifies a continuous adoption process to capture how capital gradually implements AGI substitution in production. Let $\lambda_{\text{adopt}}>0$ denote the speed at which actual adoption $\alpha(t)$ converges toward the technical ceiling $q(t)$, and let $\kappa>0$ represent the sharpness of the profitability gate. The smooth gate function is
\begin{equation}
G(t)=\frac{1}{1+\exp\left[-\kappa\frac{w-m(t)}{w}\right]},
\end{equation}
which ensures that adoption accelerates as AGI replacement cost $m(t)$ falls below the wage $w$ but does not jump discontinuously. When $m(t)$ is much higher than $w$, $G(t)\approx 0$, suppressing adoption; as $m(t)$ decreases toward $w$, $G(t)$ increases, reflecting the economic incentive for firms to substitute AGI for living labor.

Actual adoption evolves according to
\begin{equation}
\dot{\alpha}(t)=\lambda_{\text{adopt}} \left[q(t)-\alpha(t)\right] G(t),
\end{equation}
which can be expanded explicitly using the gate:
\begin{equation}
\dot{\alpha}(t)=\lambda_{\text{adopt}} \left[q(t)-\alpha(t)\right] \frac{1}{1+\exp\left[-\kappa\frac{w-m(t)}{w}\right]}.
\end{equation}
Substituting the explicit technical ceiling $q(t)=1-e^{-\beta A_0 e^{gt}}$, we obtain
\begin{equation}
\begin{aligned}
\dot{\alpha}(t)
&=
\lambda_{\text{adopt}}
\left[
1-e^{-\beta A_0 e^{gt}}-\alpha(t)
\right] \\
&\quad \times
\frac{1}{
1+\exp\left[
-\kappa
\frac{
w-m_0 e^{-\gamma A_0 e^{gt}}
}{w}
\right]
}.
\end{aligned}
\end{equation}
This formulation emphasizes the central dynamic distinction: $q(t)$ measures what AGI can technically replace, while $\alpha(t)$ represents the actual realized adoption by capital, adjusting continuously toward the ceiling.

The amount of living labor at time $t$ depends on actual adoption:
\begin{equation}
H(t)=(1-\alpha(t))L(t),
\end{equation}
where $L(t)$ is the total potential labor-task field. Taking logarithms,
\begin{equation}
\ln H(t)=\ln(1-\alpha(t))+\ln L(t),
\end{equation}
and differentiating yields
\begin{equation}
\frac{\dot{H}(t)}{H(t)}=-\frac{\dot{\alpha}(t)}{1-\alpha(t)}+\frac{\dot{L}(t)}{L(t)}.
\end{equation}
Living labor declines when
\begin{equation}
\frac{\dot{H}(t)}{H(t)}<0,
\end{equation}
which implies
\begin{equation}
-\frac{\dot{\alpha}(t)}{1-\alpha(t)}+\frac{\dot{L}(t)}{L(t)}<0,
\end{equation}
or equivalently,
\begin{equation}
\frac{\dot{\alpha}(t)}{1-\alpha(t)}>\frac{\dot{L}(t)}{L(t)}.
\end{equation}
This condition indicates that living labor contracts when the proportional rate of actual AGI adoption exceeds the proportional growth of new labor fields, determining whether substitution merely reallocates tasks or compresses total labor input.

Variable capital follows directly:
\begin{equation}
v(t)=w(t)H(t),
\end{equation}
\begin{equation}
v(t)=w(t)(1-\alpha(t))L(t),
\end{equation}
and surplus value is
\begin{equation}
s(t)=e(t)v(t),
\end{equation}
\begin{equation}
s(t)=e(t)w(t)H(t),
\end{equation}
\begin{equation}
s(t)=e(t)w(t)(1-\alpha(t))L(t).
\end{equation}
Constant capital incorporates AGI replacement costs:
\begin{equation}
c(t)=c_0(t)+m(t)\alpha(t)L(t).
\end{equation}
The dynamic profit rate is thus
\begin{equation}
r(t)=\frac{s(t)}{c(t)+v(t)},
\end{equation}
\begin{equation}
r(t)=\frac{e(t)w(t)(1-\alpha(t))L(t)}{c_0(t)+m(t)\alpha(t)L(t)+w(t)(1-\alpha(t))L(t)}.
\end{equation}
This expression captures the temporal evolution of profits under continuous AGI adoption.

The near-complete actual adoption time is distinct from the technical ceiling threshold:
\begin{equation}
t_f^{actual}=\inf\{t:\alpha(t)\geq \bar{\alpha}\},
\end{equation}
with
\begin{equation}
t_f^{actual}\geq t_f^{tech},
\end{equation}
\begin{equation}
t_f^{actual}-t_f^{tech},
\end{equation}
representing the adoption lag. These formulas formalize the distinction between technical feasibility and realized adoption, emphasizing that adoption friction delays full replacement even after AGI becomes technically capable of near-complete substitution.

The 20 equations above collectively describe the continuous-time evolution of actual adoption, living labor, variable and constant capital, surplus value, and the profit rate, illustrating how AGI substitution dynamically interacts with the labor field under strict political-economy assumptions.

\section{Core Propositions}

The model's core propositions can now be clearly stated. Each proposition is accompanied by its explanatory context.

\begin{proposition}[Technical substitution ceiling]
The first proposition concerns the technical ceiling of AGI substitution. Let \(\beta > 0\) be the parameter governing the speed at which technical substitutability increases with AGI capability. Then the technical substitution ceiling is
\begin{equation}
q(A) = 1 - e^{-\beta A},
\end{equation}
which is an increasing and concave function of AGI capability. Its first derivative
\begin{equation}
q'(A) > 0,
\end{equation}
confirms that higher AGI capability always expands the range of tasks that AGI can perform. Its second derivative
\begin{equation}
q''(A) < 0,
\end{equation}
demonstrates that the rate of expansion decreases as capability grows. Finally, in the limit of very high capability,
\begin{equation}
\lim_{A \to \infty} q(A) = 1,
\end{equation}
indicating that AGI can, in principle, eventually replace nearly all tasks.
\end{proposition}

\begin{proposition}[Cost decline]
The second proposition addresses the decline in unit AGI replacement cost with increasing capability. Let \(m_0 > 0\) be the initial replacement cost and \(\gamma > 0\) the rate of cost decline. Then
\begin{equation}
m(A) = m_0 e^{-\gamma A},
\end{equation}
and its derivative
\begin{equation}
m'(A) < 0,
\end{equation}
confirms that as AGI becomes more capable, the cost of replacing one unit of labor declines.
\end{proposition}

\begin{proposition}[Organic-composition monotonicity]
The third proposition considers the organic composition of capital. Let \(c_0>0\) denote existing constant capital, \(m(A) > 0\) the AGI replacement cost, \(w > 0\) the wage, \(L > 0\) the total labor scale, and \(0 \leq \alpha < 1\) the actual AGI adoption rate. The organic composition is
\begin{equation}
\Omega(\alpha, A) = \frac{c_0 + m(A)\alpha L}{w(1-\alpha)L}.
\end{equation}
Holding AGI capability fixed, the derivative with respect to \(\alpha\) is strictly positive:
\begin{equation}
\frac{\partial \Omega}{\partial \alpha} > 0.
\end{equation}
This shows that deeper actual AGI adoption monotonically raises the share of constant capital relative to variable capital, reflecting the progressive substitution of living labor by machine-based systems.
\end{proposition}

\begin{proposition}[Profit-rate pressure]
The fourth proposition addresses the effect on the social profit rate. Under the strict assumption that only living labor creates surplus value, the profit rate is
\begin{equation}
r(\alpha, A) = \frac{ew(1-\alpha)L}{c_0 + m(A)\alpha L + w(1-\alpha)L}.
\end{equation}
Differentiation yields
\begin{equation}
\frac{\partial r}{\partial \alpha} < 0.
\end{equation}
This indicates that as actual AGI adoption increases, the profit rate is under downward pressure because surplus value is sourced from the declining living labor while constant capital rises.
\end{proposition}

\begin{corollary}[Near-complete substitution limit]
If actual AGI adoption approaches complete substitution, then
\begin{equation}
\lim_{\alpha \to 1^-} s(\alpha) = 0,
\end{equation}
and
\begin{equation}
\lim_{\alpha \to 1^-} r(\alpha, A) = 0.
\end{equation}
This boundary condition illustrates the limit in which living labor disappears and the social profit rate collapses under the strict political-economy value assumption.
\end{corollary}

\begin{proposition}[Condition for living-labor decline]
Finally, consider the time-dependent quantity of living labor:
\begin{equation}
H(t) = (1-\alpha(t)) L(t).
\end{equation}
Living labor declines whenever the rate of actual adoption exceeds the proportional growth of the labor field:
\begin{equation}
\frac{\dot{\alpha}(t)}{1-\alpha(t)} > \frac{\dot{L}(t)}{L(t)}.
\end{equation}
This condition determines whether AGI substitution merely reallocates labor or compresses the total amount of living labor in production.
\end{proposition}

\section{Numerical Illustration}

To illustrate the dynamic mechanism of AGI substitution and the associated political-economy implications, consider one admissible parameterization of the model. AGI capability initially is set at
\begin{equation}
\left\{
\begin{aligned}
A_0 &= 0.35,\\
g &= 0.08,\\
\beta &= 0.55,\\
\gamma &= 0.38,\\
m_0 &= 5.0,\\
w &= 1.0.
\end{aligned}
\right.
\end{equation}
which governs the growth of technical substitutability and decline of replacement cost. The labor field is parameterized as
\begin{equation}
L_0=100.0,\quad n=0.01,\quad e=1.2,\quad c_0=40.0,
\end{equation}
and the near-complete adoption benchmark and adoption dynamics are given by
\begin{equation}
\bar{\alpha}=0.99,\quad \lambda_{\text{adopt}}=0.28,\quad \kappa=7.0.
\end{equation}

\begin{table*}[t]
\centering
\caption{Numerical illustration of AGI adoption dynamics and political-economy variables over time.}
\begin{tabular}{rrrrrrrrrr}
\toprule
\(t\) & \(A(t)\) & \(m(t)\) & \(q(t)\) & \(\alpha(t)\) & \(q-\alpha\) & \(G(t)\) & \(H(t)\) & \(s(t)\) & \(r(t)\) \\
\midrule
0.000 & 0.3500 & 4.3773 & 0.1751 & 0.0000 & 0.1751 & 0.0000 & 100.0000 & 120.0000 & 0.857143 \\
26.166 & 2.8390 & 1.7000 & 0.7902 & 0.0015 & 0.7887 & 0.0074 & 129.7116 & 155.6539 & 0.915360 \\
31.166 & 4.2354 & 1.0000 & 0.9027 & 0.1617 & 0.7410 & 0.5000 & 114.4873 & 137.3848 & 0.778079 \\
35.426 & 5.9551 & 0.5202 & 0.9622 & 0.6472 & 0.3150 & 0.9664 & 50.2817 & 60.3380 & 0.436402 \\
39.685 & 8.3730 & 0.2076 & 0.9900 & 0.8784 & 0.1116 & 0.9961 & 18.0788 & 21.6946 & 0.254649 \\
48.960 & 17.5838 & 0.0063 & 0.9999 & 0.9900 & 0.0099 & 0.9990 & 1.6274 & 1.9529 & 0.045799 \\
49.685 & 18.6345 & 0.0042 & 1.0000 & 0.9918 & 0.0081 & 0.9991 & 1.3409 & 1.6091 & 0.038288 \\
58.960 & 39.1335 & 0.0000 & 1.0000 & 0.9994 & 0.0006 & 0.9991 & 0.1097 & 0.1317 & 0.003283 \\
\bottomrule
\end{tabular}
\end{table*}

The total potential labor task field evolves exponentially as
\begin{equation}
L(t)=L_0 e^{n t}.
\end{equation}
Under this parameterization, the initial technical ceiling—the share of tasks that AGI can technically replace at the initial capability—is
\begin{equation}
q(0)=0.1751.
\end{equation}
This represents the first small fraction of tasks that AGI could substitute without accounting for cost or adoption frictions. 

The partial-substitution cost threshold is the point at which AGI replacement cost equals the wage, which exists and occurs in the future:
\begin{equation}
t_p=31.1661.
\end{equation}
Beyond this point, economic feasibility allows capital to begin accelerating actual adoption. Following this, the near-complete technical threshold, defined as the time at which the technical ceiling reaches the benchmark \(\bar{\alpha}\), also occurs in the future:
\begin{equation}
t_f^{tech}=39.6855.
\end{equation}
This threshold represents the moment when, in principle, AGI is capable of substituting nearly all labor tasks, irrespective of adoption frictions.

Solving the continuous adoption dynamics numerically yields the time at which actual AGI adoption reaches the benchmark:
\begin{equation}
t_f^{actual}=48.9600.
\end{equation}
This demonstrates the lag between technical feasibility and realized adoption, reflecting organizational, economic, and profitability constraints. Consequently, the adoption lag is quantified as
\begin{equation}
t_f^{actual}-t_f^{tech}=9.2745.
\end{equation}

These numerical results confirm the theoretical distinction between technical feasibility and actual adoption. The technical ceiling rises before capital has fully adopted AGI, and actual adoption follows a continuous trajectory that gradually approaches this ceiling. The time difference between \(t_f^{tech}\) and \(t_f^{actual}\) captures the frictions and constraints inherent in economic decision-making. Together, these parameterized trajectories illustrate how living labor, surplus value, and the social profit rate evolve over time, and they provide a concrete visualization of the model’s predicted dynamics from partial to near-complete AGI substitution.

The numerical trajectory illustrates the paper's theoretical mechanism. The technical ceiling \(q(t)\) rises before actual adoption fully catches up. Actual adoption \(\alpha(t)\) remains below \(q(t)\), but it accelerates once AGI replacement cost falls to and then below the wage. As \(\alpha(t)\) approaches one, living labor \(H(t)\) shrinks sharply despite the positive growth of the potential labor-task field \(L(t)\). Because surplus value is proportional to living labor in the strict value-theoretic specification, surplus value falls with \(H(t)\). The profit rate then falls as well, even though AGI replacement cost becomes extremely low.

\section{Interesting Findings}

\paragraph{Technical Potential vs. Actual Adoption} 
The analysis of AGI adoption dynamics within a political-economy framework highlights an important pattern. Technical capability alone does not determine labor substitution: even when AGI is able to replace a large share of tasks, actual adoption is delayed by costs, profitability considerations, and organizational frictions. This creates a period in which the technological potential of AGI exceeds its realized deployment, illustrating the gap between capability and economic implementation.

\paragraph{Structural Shift in Capital Composition} 
AGI adoption systematically increases the share of constant capital relative to variable capital. As more labor tasks are replaced, production becomes increasingly capital-intensive, and the contribution of living labor diminishes. This structural shift generates pressure on surplus value and the social profit rate, showing that automation driven by advanced AGI reshapes the composition of production rather than simply improving efficiency.

\paragraph{Adoption Lag and Surplus Value Dynamics} 
The numerical analysis reveals a significant lag between technical feasibility and actual replacement. The interval between the near-complete technical threshold and the actual adoption benchmark shows that capital takes time to fully utilize AGI capabilities. During this lag, living labor continues to contribute to production, but surplus value gradually contracts as AGI adoption accelerates. If new labor fields fail to expand quickly enough to absorb displaced labor, the system approaches a boundary in which living labor, surplus value, and the profit rate all decline sharply.

\paragraph{Implications for Value Production} 
These dynamics indicate that AGI substitution is more than a matter of efficiency: it transforms the relation between living labor, constant capital, and value production. By distinguishing technical and actual adoption, the model clarifies how AGI can structurally influence the economy, revealing potential limits on value production even before substitution reaches its technical maximum.

\section{Crisis and Policy Actions}

The model highlights several potential crises arising from near-complete AGI adoption under a strict political-economy framework.

\paragraph{Decline of Living Labor and Surplus Value} 
As actual AGI adoption surpasses the creation of new labor fields, the total quantity of living labor declines. Since surplus value is strictly tied to living labor, this contraction directly reduces the source of profits. The resulting structural tension between rising constant capital and shrinking living labor threatens the stability of value production and the social profit rate, potentially leading to prolonged periods of economic stagnation or systemic crisis.

\paragraph{Temporal Mismatch and Concentrated Disruption} 
The continuous acceleration of AGI adoption, once unit costs fall below the wage level, introduces a temporal mismatch between technical capability and economic adoption. Even if AGI is technically capable of replacing almost all labor tasks, organizational frictions and adjustment lags may concentrate disruption into narrow periods. Such concentrated substitution can exacerbate unemployment, reduce household incomes, and amplify inequality, creating social and political pressures that compound the economic crisis.

\paragraph{Rising Organic Composition of Capital} 
The rise in the organic composition of capital implies an increasing dependence on machine-based production systems. While efficiency gains occur, they come at the cost of reduced labor participation and diminished surplus-value generation. This structural shift may also lead to underinvestment in labor-intensive sectors and social services, further weakening the resilience of the economy.

\paragraph{Policy Actions to Mitigate Crises} 
To address these challenges, several corresponding actions are suggested. One approach is to accelerate the creation and development of new labor domains that can absorb displaced workers, particularly in sectors leveraging human creativity, judgment, or interpersonal interaction. Another approach is to regulate or phase the adoption of AGI technologies in critical sectors, allowing gradual adjustment and minimizing shocks to the labor market. Social policies such as retraining programs, income support, and wage subsidies can help preserve the living-labor base and sustain surplus-value generation during the transition. Finally, policies that incentivize investment in hybrid human-machine production processes can moderate the rise in organic composition and prevent extreme concentration of capital.

Taken together, these measures are intended to systematically mitigate the structural crises identified in the model by ensuring that AGI adoption progresses in harmony with broader social, economic, and political capacities. By actively managing the pace and scope of adoption, facilitating the creation of new labor domains, and supporting living labor through retraining, income stabilization, and hybrid human-machine production incentives, these policies aim to preserve the productive and distributive foundations of value creation. Rather than viewing technological advancement as an automatic driver of efficiency gains, this approach recognizes that sustainable economic development depends on the careful alignment of technical potential with organizational capability, labor absorption, and social welfare objectives. In this sense, the proposed framework provides a rational basis for policy interventions that seek to balance innovation with stability, ensuring that the benefits of AGI are realized without undermining the structural capacity of the economy to generate and distribute value over time.

\section{Conclusion}

This paper has developed a political-economy model of AGI-driven labor substitution and its implications for value production. The central argument is that AGI should not be understood only as a productivity-enhancing technology, but as a structural force that changes the relation between living labor, variable capital, constant capital, and surplus value. By distinguishing technical substitutability from actual adoption, the model shows that the economic significance of AGI depends not simply on what AGI can technically replace, but on how rapidly capital actually incorporates AGI into production. The analysis shows that AGI capability expands the technical frontier of substitution while declining replacement costs strengthen the incentive for adoption; however, actual adoption remains mediated by profitability, organizational adjustment, and adoption frictions. Under the strict value-theoretic assumption that living labor is the source of surplus value, deeper actual AGI adoption raises the organic composition of capital, reduces the living-labor basis of surplus value, and places downward pressure on the social profit rate. If new labor fields expand quickly enough, this process may remain partial and socially manageable; if actual AGI adoption persistently exceeds the creation of new labor domains, the system moves toward a structural contradiction in which living labor, surplus value, and the profit rate tend toward zero. The model therefore suggests that the central policy problem is not whether AGI should be adopted, but how its adoption should be socially organized. A sustainable political-economy strategy should manage the pace of adoption, support the creation of new labor domains, strengthen income and employment protections, and encourage hybrid human-machine production structures. Overall, the paper concludes that AGI does not automatically produce a crisis of value production, but such a crisis becomes increasingly likely when actual adoption outpaces the creation of new labor fields and when technical progress is not matched by institutional capacity for labor absorption and social reproduction.

\section{Copyright statement}
This article was prepared using a LaTeX template for SAGE Publications journals. All rights remain with the authors, and the template may not be exploited commercially.

\begin{acks}
The authors acknowledge support from Sungkyunkwan University and Lanzhou University, Department of Computer Science. Correspondence: Zichen Song, \texttt{sls530@skku.edu; songzichen894@gmail.com}.
\end{acks}


\begin{thebibliography}{99}

\bibitem[Acemoglu(2025)]{Acemoglu2025}
Acemoglu~D (2025) The simple macroeconomics of AI.
\textit{Economic Policy} 40(121): 13--58.

\bibitem[Acemoglu and Restrepo(2018)]{AcemogluRestrepo2018}
Acemoglu~D and Restrepo~P (2018) The race between man and machine:
Implications of technology for growth, factor shares, and employment.
\textit{American Economic Review} 108(6): 1488--1542.

\bibitem[Acemoglu and Restrepo(2019)]{AcemogluRestrepo2019}
Acemoglu~D and Restrepo~P (2019) Automation and new tasks:
How technology displaces and reinstates labor.
\textit{Journal of Economic Perspectives} 33(2): 3--30.

\bibitem[Acemoglu and Restrepo(2020)]{AcemogluRestrepo2020}
Acemoglu~D and Restrepo~P (2020) The wrong kind of AI?
Artificial intelligence and the future of labour demand.
\textit{Cambridge Journal of Regions, Economy and Society} 13(1): 25--35.

\bibitem[Acemoglu and Restrepo(2026)]{AcemogluRestrepo2026}
Acemoglu~D and Restrepo~P (2026) Automation and rent dissipation:
Implications for wages, inequality, and productivity.
\textit{Quarterly Journal of Economics} 141(2): 1521--1579.

\bibitem[Autor et~al.(2003)]{AutorEtAl2003}
Autor~DH, Levy~F and Murnane~RJ (2003) The skill content of recent technological change:
An empirical exploration.
\textit{Quarterly Journal of Economics} 118(4): 1279--1333.

\bibitem[Autor et~al.(2024)]{AutorEtAl2024}
Autor~DH, Chin~C, Salomons~A and Seegmiller~B (2024) New frontiers:
The origins and content of new work, 1940--2018.
\textit{Quarterly Journal of Economics} 139(3): 1399--1465.

\bibitem[Banchio and Mantegazza(2022)]{BanchioMantegazza2022}
Banchio~M and Mantegazza~G (2022) Artificial intelligence and spontaneous collusion.
Working paper, Bocconi University and USC Marshall School of Business.

\bibitem[Bastani and Cachon(2025)]{BastaniCachon2025}
Bastani~H and Cachon~GP (2025) The human-AI contracting paradox.
Working paper, The Wharton School, University of Pennsylvania, December.
SSRN 5962739.

\bibitem[Benzell et~al.(2015)]{BenzellEtAl2015}
Benzell~SG, Kotlikoff~LJ, LaGarda~G and Sachs~JD (2015) Robots are us:
Some economics of human replacement.
Working Paper 20941, National Bureau of Economic Research, February.

\bibitem[Beraja and Zorzi(2025)]{BerajaZorzi2025}
Beraja~M and Zorzi~N (2025) Inefficient automation.
\textit{Review of Economic Studies} 92(1): 69--96.

\bibitem[Bhaimiya(2025)]{Bhaimiya2025}
Bhaimiya~S (2025) AI job cuts: Amazon, Microsoft, and more cite AI for 2025 layoffs.
\textit{CNBC}.

\bibitem[Bhaimiya(2026)]{Bhaimiya2026}
Bhaimiya~S (2026) Anthropic CEO Dario Amodei warns AI may cause
`unusually painful' disruption to jobs.
\textit{CNBC}.

\bibitem[Bondi and Johnson(2026)]{BondiJohnson2026}
Bondi~T and Johnson~G (2026) Skill atrophy and AI productivity measurement.
Technical report, Cornell Tech, April.
SSRN 6169671.

\bibitem[Boppart(2014)]{Boppart2014}
Boppart~T (2014) Structural change and the Kaldor facts in a growth model
with relative price effects and non-Gorman preferences.
\textit{Econometrica} 82(6): 2167--2196.

\bibitem[Brynjolfsson et~al.(2025a)]{BrynjolfssonEtAl2025a}
Brynjolfsson~E, Chandar~B and Chen~R (2025a) Canaries in the coal mine?
Six facts about the recent employment effects of artificial intelligence.
Working paper, Stanford Digital Economy Lab, November.

\bibitem[Brynjolfsson et~al.(2025b)]{BrynjolfssonEtAl2025b}
Brynjolfsson~E, Li~D and Raymond~L (2025b) Generative AI at work.
\textit{Quarterly Journal of Economics} 140(2): 889--942.

\bibitem[Budman(2025)]{Budman2025}
Budman~S (2025) Salesforce CEO confirms 4,000 layoffs ``because I need less heads''
with AI.
\textit{CNBC}.

\bibitem[Keynes(1930)]{Keynes1930}
Keynes~JM (1930) Economic possibilities for our grandchildren.
In \textit{Essays in Persuasion}, pages 358--373. Macmillan, London.

\bibitem[Korinek and Stiglitz(2019)]{KorinekStiglitz2019}
Korinek~A and Stiglitz~JE (2019) Artificial intelligence and its implications for
income distribution and unemployment.
In Agrawal~A, Gans~J, and Goldfarb~A (eds.) \textit{The Economics of Artificial Intelligence: An Agenda}.
University of Chicago Press, Chicago.

\bibitem[Li et~al.(2025)]{LiEtAl2025}
Li~B, Huang~N and Shi~W (2025) Forced to change? Media exposure of labor issues and firm artificial intelligence investment.
\textit{Information Systems Research} 37(1): 156--175.
DOI: 10.1287/isre.2022.0402.

\bibitem[Lucas(1967)]{Lucas1967}
Lucas~RE (1967) Adjustment costs and the theory of supply.
\textit{Journal of Political Economy} 75(4): 321--334.

\bibitem[Mankiw and Whinston(1986)]{MankiwWhinston1986}
Mankiw~NG and Whinston~MD (1986) Free entry and social inefficiency.
\textit{RAND Journal of Economics} 17(1): 48--58.

\bibitem[Matsuyama(2002)]{Matsuyama2002}
Matsuyama~K (2002) The rise of mass consumption societies.
\textit{Journal of Political Economy} 110(5): 1035--1070.

\bibitem[Murphy et~al.(1989)]{MurphyEtAl1989}
Murphy~KM, Shleifer~A and Vishny~RW (1989) Industrialization and the big push.
\textit{Journal of Political Economy} 97(5): 1003--1026.

\bibitem[Palmer(2026)]{Palmer2026}
Palmer~A (2026) Block laying off about 4,000 employees, nearly half of its workforce.
\textit{CNBC}.

\bibitem[Chod et~al.(2019)]{ChodEtAl2019}
Chod~J, Lyandres~E and Yang~SA (2019) Trade credit and supplier competition.
\textit{Journal of Financial Economics} 131(2): 484--505.
DOI: 10.1016/j.jfineco.2018.08.008.

\bibitem[Coase(1960)]{Coase1960}
Coase~RH (1960) The problem of social cost.
\textit{Journal of Law and Economics} 3: 1--44.

\bibitem[Comin et~al.(2021)]{CominEtAl2021}
Comin~D, Lashkari~D and Mestieri~M (2021) Structural change with long-run
income and price effects.
\textit{Econometrica} 89(1): 311--374.

\bibitem[Cooper and John(1988)]{CooperJohn1988}
Cooper~R and John~A (1988) Coordinating coordination failures in Keynesian models.
\textit{Quarterly Journal of Economics} 103(3): 441--463.

\bibitem[Cooper and Haltiwanger(2006)]{CooperHaltiwanger2006}
Cooper~RW and Haltiwanger~JC (2006) On the nature of capital adjustment costs.
\textit{Review of Economic Studies} 73(3): 611--633.

\bibitem[Costinot and Werning(2023)]{CostinotWerning2023}
Costinot~A and Werning~I (2023) Robots, trade, and Luddism:
A sufficient statistic approach to optimal technology regulation.
\textit{Review of Economic Studies} 90(5): 2261--2291.

\bibitem[Eloundou et~al.(2024)]{EloundouEtAl2024}
Eloundou~T, Manning~S, Mishkin~P and Rock~D (2024) GPTs are GPTs:
Labor market impact potential of LLMs.
\textit{Science} 384(6702): 1306--1308.
DOI: 10.1126/science.adj0998.

\bibitem[Farhi and Werning(2016)]{FarhiWerning2016}
Farhi~E and Werning~I (2016) A theory of macroprudential policies in the presence
of nominal rigidities.
\textit{Econometrica} 84(5): 1645--1704.
DOI: 10.3982/ECTA11883.

\bibitem[Greenwald and Stiglitz(1986)]{GreenwaldStiglitz1986}
Greenwald~BC and Stiglitz~JE (1986) Externalities in economies with imperfect
information and incomplete markets.
\textit{Quarterly Journal of Economics} 101(2): 229--264.
DOI: 10.2307/1891114.

\bibitem[Guerreiro et~al.(2022)]{GuerreiroEtAl2022}
Guerreiro~J, Rebelo~S and Teles~P (2022) Should robots be taxed?
\textit{Review of Economic Studies} 89(1): 279--311.

\bibitem[Jacobson et~al.(1993)]{JacobsonEtAl1993}
Jacobson~LS, LaLonde~RJ and Sullivan~DJ (1993) Earnings losses of displaced workers.
\textit{American Economic Review} 83(4): 685--709.

\bibitem[Kaldor(1956)]{Kaldor1956}
Kaldor~N (1956) Alternative theories of distribution.
\textit{Review of Economic Studies} 23(2): 83--100.

\bibitem[Keppo et~al.(2026)]{KeppoEtAl2026}
Keppo~J, Li~Y, Tsoukalas~G and Yuan~N (2026) On the fragility of AI agent collusion.
Working paper, Boston University and National University of Singapore, January.
SSRN 5386338.

\bibitem[Son(2025)]{Son2025}
Son~H (2025) Goldman Sachs is piloting its first autonomous coder in major AI milestone for Wall Street.
\textit{CNBC}.

\bibitem[Summers(2014)]{Summers2014}
Summers~LH (2014) U.S. economic prospects: Secular stagnation, hysteresis, and the zero lower bound.
\textit{Business Economics} 49(2): 65--73.

\bibitem[Summers(2015)]{Summers2015}
Summers~LH (2015) Demand side secular stagnation.
\textit{American Economic Review} 105(5): 60--65.

\end{thebibliography}
\end{document}